\documentclass[article,aps]{revtex4}
\usepackage{dcolumn}
\usepackage{bm}
\usepackage{amsmath}
\usepackage{amsfonts}
\usepackage{amssymb}
\usepackage{graphicx}
\usepackage{color}
\newcommand{\beq}{\begin{eqnarray}}
\newcommand{\eeq}{\end{eqnarray}}
\newcommand{\slp}{\raise.15ex\hbox{$/$}\kern-.57em\hbox{$ \partial $}}
\newcommand{\lnA}{\raise.15ex\hbox{$/$}\kern-.65em\hbox{$A$}}
\newcommand{\lns}{\raise.15ex\hbox{$/$}\kern-.57em\hbox{$s$}}
\newcommand{\lnb}{\raise.15ex\hbox{$/$}\kern-.57em\hbox{$b$}}
\renewcommand{\vec}[1]{{\mathbf{#1}}}

\providecommand{\abs}[1]{\lvert#1\rvert}
\begin{document}

\title{Duality and  bosonization in Schwinger-Keldysh formulation}
\author{R. E. Gamboa Sarav\'i}
\author{C. M. Na\'on}
\author{F. A. Schaposnik}
\affiliation{Departamento de F\'{\i}sica, Facultad de Ciencias
Exactas, Universidad Nacional de La Plata and IFLP-CONICET, CC 67,
 1900 La Plata, Argentina.}
\begin{abstract}
We present a path-integral bosonization approach for systems out of equilibrium  based on a duality transformation of the original Dirac fermion theory combined with the Schwinger-Keldysh time closed contour technique, to handle the non-equilibrium situation. The duality approach to bosonization that we present is valid for $D \geq 2$ space-time dimensions leading for $D=2$ to exact results.   In this last case we present the  bosonization rules    for fermion currents, calculate current-current correlation functions and establish the connection  between the fermionic and bosonic distribution functions in a generic, nonequilibrium situation.
\end{abstract}
\date{\today}
 \maketitle

\section{Introduction}
Bosonization is a powerful technique, widely used to analyze quantum systems in one spatial dimension \cite{Stone}. In the context of relativistic quantum field theories (QFT), following the seminal papers by Coleman, Mandelstam and Witten
\cite{Coleman}-\cite{Witten}, the bosonization procedure became a key tool in the solution of simplified models of strong interactions, such as QCD in $(1+1)$ dimensions. These developments also gave strength to the idea of duality, which is currently exploited with big success in the framework of string theories \cite{strings}.  {On the other hand, bosonization in condensed matter physics \cite{M1}-\cite{M2} is an extremely important technique in the study of non-relativistic, one-dimensional systems (see
 \cite{reviews} and reference therein)}. Indeed, the Luttinger liquid state, which manifests experimentally in carbon nanotubes, polymer nanowires and quantum Hall edges, can be naturally understood in terms of bosonic collective modes, obtained analytically through bosonization. It is important to emphasize that  the bosonization method was restricted to equilibrium physical situations until recently, when Gutman, Gefen and Mirlin  in a significant work  \cite{GGM} built a bosonic theory for 1D fermions under non-equilibrium conditions. Starting from a model of free fermions of a definite chirality, they obtained an equivalent bosonic effective action containing terms of all orders in the fields.  It should be stressed that this is an important achievement in view of the current, growing interest in non-equilibrium phenomena in nanoscopic systems \cite{nano-exp}.

In this work we present an out of equilibrium path-integral bosonization approach based in duality transformations  proposed in \cite{Betal}-\cite{modern-bos}. Our proposal is inspired in the "target space duality" of string theory \cite{GPR}, and can be applied in any number of space dimensions, leading to exact results for Abelian and non-Abelian bosonization of massive and massless Dirac fermions in $1+1$ space-time dimension and perturbative answers in $d>1$ space dimensions. We find that this path-integral bosonization framework is particularly appropriate when considering out of equilibrium systems using the Schwinger-Keldysh time closed contour technique \cite{Schwinger}-\cite{Das} leading to very natural extensions
 of the Coleman-Mandelstam bosonization recipe \cite{Coleman}.

 The paper is organized as follows. In section II we introduce the path-integral generating functional for   non-interacting   Dirac fermions with dynamics governed
by an action defined on the Schwinger-Keldish time contour. We discuss boundary conditions in different branches of the contour and introduce the retarded, advanced and Keldish fermionic Green functions. Then, in Section III, we present the path-integral bosonization approach based in duality transformations. After gauging the global symmetry of the original fermionic theory and introducing a Lagrange multiplier that constraints the auxiliary gauge field to be a pure gauge, we arrive to a bosonic dual and establish the correspondence between fermionic and bosonic currents. In Section IV we present the calculation of current-current correlation functions both in the bosonic and fermionic sectors and establish the connection  between the fermionic and bosonic distribution
functions   for a generic, nonequilibrium situation. We summarize and discuss our results in section V, where we also comment  possible extensions of our approach to $D>2$ space-time dimensions.

\section{Generating functional for free Dirac fermions in the Schwinger-Keldysh time closed path \label{section2}}
\subsection{The model}
We start by considering a system of fermion fields $\psi$ and $\psi^{\dagger}$ in $D=d+1$ space-time dimensions (later on we will specialize our results to the case $d=1$). The Lagrangian density is given by
\beq
{\cal L}_{F}=\bar{\psi}i \slp \psi
 \eeq
Here $\slp \equiv \gamma^{\mu} \partial_{\mu}$, with $\gamma^{\mu}$  the Dirac matrices and  { $(x^\mu) \equiv (x^0= t,\vec x) $}.

\begin{figure}
\centering
\includegraphics[scale=.7]{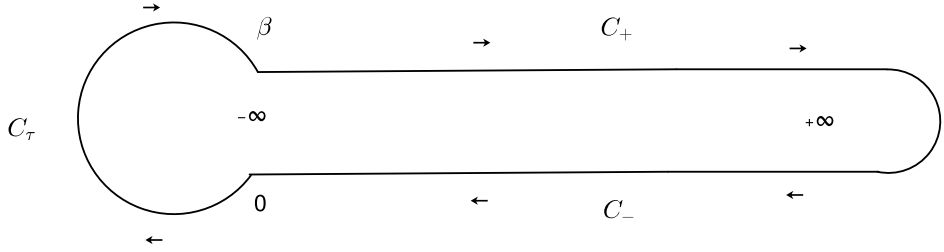}
\caption{(Color online) Sketch of the closed time path contour. The upper branch of the path, $C_+$, represents the usual time path,
from $t=-\infty$ to $t=+\infty$. The lower branch $C_-$ goes backward in time. Both branches are connected, in the remote past, by the curve $C_{\tau}$.}
\label{fig1}
\end{figure}

As it is well-known, the Schwinger-Keldysh formulation \cite{Schwinger}-\cite{Keldysh}
 enables to extend the validity of a quantum field theory to the realm of non-equilibrium physics by
defining the action $S[\psi,\bar{\psi},s_{\mu}]$ along a closed time contour ${\cal C}=C_+ \cup C_- \cup C_{\tau}$ \cite{Babichenko-Kozlov} represented in Fig.\ref{fig1}. Within this formulation,
we
 start by considering the partition function
\begin{equation}\label{partition}
Z_F= \int\,{\cal D}\bar{\psi}\,{\cal D}\psi\,e^{i S_F[\psi,\bar{\psi}]},
\end{equation}
with
\begin{equation}
S_F[\psi,\bar\psi]
=\int_{\cal C}\,dt\,d^{d} x \, \bar\psi i \slp \psi,
\label{principio}
\end{equation}
where subscript ${\cal C}$ indicates that   the action is defined along a closed time path \cite{Babichenko-Kozlov} so that in the path-integral defining $Z_F$ the fermion fields should be integrated imposing  the appropriate boundary conditions
on each branch of the contour.

As indicated in Fig.1,  the contour is splitted in two branches, one from $t=-\infty$ to $t=+\infty$ ($C_+$),
the other one completing the contour ($C_-$). In the remote past ($t=-\infty$) the forward and backward paths are connected by $C_{\tau}$ {where $\tau \in (0,\beta)$ is imaginary time when evolving along ${\cal C}_\tau$ and
  $\beta = 1/T$ is the inverse temperature.} The precise form of this connection depends on the initial distribution, which can be assumed to be an equilibrium distribution.
Fields in each one of the two contour sections  $C_+$ and $C_-$ will be denoted as $\psi_+$, $\psi_+^{\dagger}$,  and $\psi_-$, $\psi_-^{\dagger}$, respectively. Concerning fields in $C_{\tau}$, they will be called $\psi_{\tau}$, $\psi_{\tau}^{\dagger}$.
With this, action $S_F$ will be written in the form
 \begin{equation}\label{action}
S_F[\psi,\bar{\psi} ]= \int d^d x  \int_{-\infty}^\infty dt \left({\cal L}[\psi_+, \bar{\psi}_{+}]-{\cal L}[\psi_-, \bar{\psi}_{-}]\right) +  \int_{\cal C_\tau}  dt d^dx {\cal L_\tau}
\end{equation}
where
\begin{eqnarray}
{\cal L}[\psi_{\pm}, \bar{\psi}_{\pm}] &=& \bar{\psi}_{\pm} i \slp \psi_{\pm}. \label{novi1}
\\
{\cal L}_\tau[\psi_\tau, \bar{\psi}_\tau] &=& \bar{\psi}_\tau i \slp {\psi_\tau}
\end{eqnarray}

 {The appropriate boundary conditions for the fermion fields are \cite{Babichenko-Kozlov}
\begin{eqnarray}
\psi_{+} (\vec x,t =+\infty) &=& \psi_{-}(\vec x, t= +\infty)\nonumber\\
\psi_{+}^{\dagger} (\vec x,t =+\infty) &=& \psi_{-}^{\dagger}(\vec x, t= +\infty)\nonumber\\
\psi_{-} (\vec x,t =-\infty) &=& \psi_{\tau}(\vec x, t= 0)\nonumber\\
\psi_{-}^{\dagger} (\vec x,t =-\infty) &=& \psi_{\tau}^{\dagger}(\vec x, t= 0)\nonumber\\
\psi_{+} (\vec x,t =-\infty) &=& -\psi_{\tau}(\vec x, t= \beta)\nonumber\\
\psi_{+}^{\dagger} (\vec x,t =-\infty) &=& -\psi_{\tau}^{\dagger}(\vec x, t= \beta),
\label{boundary}
\end{eqnarray}}
where the minus sign in the two last equations is related to the Grassmanian nature of fermionic variables. It is through   these boundary conditions that the fermion fields $\psi_\pm, \psi_\tau$ are coupled. The role of the field $\psi_\tau$ is precisely  to
provide matching conditions for $\psi_\pm$ and they will play no role in  final expressions  \cite{KO}. That is the reason why when we define the generating functional, in order to compute vacuum expectation values,
we do not couple sources to $\psi_\tau$ fields and write
\beq
Z_F[s_\mu] = \int {\cal D}\bar\psi_+ {\cal D}\bar\psi_-{\cal D}\psi_+ {\cal D}\psi_- \exp\left(i S_F[\bar \psi,  \psi, s ] \right)
\label{zita}
\eeq
where
\beq
S_F[\bar \psi,  \psi, s ]
 =
 \int d^d x  \int_{-\infty}^\infty dt \left({\cal L}[\psi_+, \bar{\psi}_{+}, s_{\mu}^+]-{\cal L}[\psi_-, \bar{\psi}_{-}, s_{\mu}^-]\right)
\eeq
Written in this form, one could think that the partition function \eqref{zita} consists of two independent contributions which can be treated separately so that the partition function splits in two independent factors.  However, as indicated above, the path-integration requires boundary conditions which in the present case  are precisely those in eq.\eqref{boundary}  which do couple the $\pm$-fields.

\subsection{Free fermions in the time contour}

 Let us now  show  how to take into account, within the path-integral formulation,  the time boundary conditions.
Let us denote   $\psi_a$ and $\bar{\psi}_a$  the spinors introduced in \eqref{action},    with $a=+,-$. Thus, the free fermion action can be written in terms of the time contour matrix structure as

\begin{equation}\label{free-action}
S_{F}[\psi,\bar{\psi},0]=\int d^dx\,dt\,\bar{\psi}_a\,D_{ab}\,\psi_b,
\end{equation}
where
\[D=\left(\!\!\begin{array}{cc}i \slp\! & 0\\
0\! & -i \slp\\
\end{array}\!\!\right).\]
As discussed above  the required boundary conditions \eqref{boundary}   imply a correlation between $+$ and $-$ fields at the boundaries so that although $D^0$ is diagonal, its inverse operator (the Green function) {could have non vanishing off-diagonal components}. Indeed, the time boundary terms lead to a Green function which is a nondiagonal matrix
\[G=\left(\!\!\begin{array}{cc}G_{++}\! & G_{+-}\\
G_{-+}\! & G_{--}\\
\end{array}\!\!\right)\label{green-libre},\]
satisfying
\[D_x\, G_{xy}=\delta(t_x-t_y) \delta^{d}(\vec x-\vec y)   I_{4 \times 4} \, .\]
or, more explicitly, for the case of one spatial dimension
\beq
i\left(
\begin{array}{cccc}
0 & \partial_0 + \partial_1 & 0 & 0\\
\partial_0 - \partial_1 & 0 & 0 & 0\\
0 & 0 & 0 & -(\partial_0 + \partial_1)\\
0 & 0& -(\partial_0 - \partial_1) & 0\end{array}
\right)
\left(
\begin{array}{cccc}
0 & G_{++}^R & 0 & G_{+-}^R\\
G_{++}^L & 0 & G_{+-}^L & 0\\
0 & G_{-+}^R & 0 & G_{--}^R\\
G_{-+}^L & 0& G_{--}^L& 0\end{array}
\right)
= \delta(t_x-t_y)   \delta( x-  y)   I_{4 \times 4},
\label{restoran}
\eeq
%
where the superscripts $R$ and $L$ indicate the chiral components of the fermionic Green functions.

Note that when considering the usual equilibrium situation, the introduction of the contour is not mandatory and only the time forward branch is usually taken into account. Consequently, the Green function is just $G_{++}$, which is nothing but the time-ordered propagator. Let us mention, however, that in some applications the use of the closed time contour can be convenient even in equilibrium situations. In the present formulation $G_{--}$ represents an anti-time-ordered propagator. The off-diagonal components $G_{+-}$ and $G_{-+}$ satisfy homogeneous differential equations and depend on the distribution function. Moreover, it can be shown that only three of the four elements of $G$ are independent. Indeed, one can verify that $G_{++}+G_{--}=G_{+-}+G_{-+}$. This suggests that the fermionic Green function can be written as a triangular matrix. In fact, following Larkin and Ovchinnikov \cite{Larkin-Ovchinnikov}, we perform the following rotation of fermion fields:
\beq\Psi_1=\frac{\psi_+ + \psi_-}{\sqrt{2}}\,\,\,\,\,\,\,\,\,\,\,\,\,\,\,\,\,\,  \Psi_2=\frac{\psi_+ - \psi_-}{\sqrt{2}} \!\!\eeq
and
\beq\bar{\Psi}_1=\frac{\bar{\psi}_{+} - \bar{\psi}_{-}}{\sqrt{2}}\,\,\,\,\,\,\,\,\,\,\,\,\,\,\,\,\,\, \bar{\Psi}_2=\frac{\bar{\psi}_{+} + \bar{\psi}_{-}}{\sqrt{2}}  \!\! .\eeq
These changes can be implemented introducing   two matrices $M$ and $N$, such that
\beq\label{LO}\psi=M \Psi\,\,\,\,\,\,\,\,\,\,\,\,\,\,\,\,\,\,  \bar{\psi}=\bar{\Psi}N, \!\!\eeq
where $\Psi$ and $\bar{\Psi}$ {are now spinors constructed from $\Psi_1, \Psi_2$}, and
\[M=\frac{1}{\sqrt{2}}\left(\!\!\begin{array}{cc}1\! & 1\\
1\! & -1\\
\end{array}\!\!\right),\;\;\;\;\;N=\frac{1}{\sqrt{2}}\left(\!\!\begin{array}{cc}1\! &-1\\
1\! & 1\\
\end{array}\!\!\right).\]
In terms of these new fields the free action becomes
\begin{equation}\label{free-action2}
S_{F}[\Psi,\bar{\Psi},0]=\int dx\,dt\,\bar{\Psi}\,{\cal D} \,\Psi,
\end{equation}
where now the free Dirac operator reads
\beq\label{D0}{\cal D}^0 =\left(\!\!\begin{array}{cc}i \slp\! & 0\\
0\! & i \slp\\
\end{array}\!\!\right).\eeq
 The boundary conditions for spinors can be easily determined from those imposed to $\psi_{\pm}$ and $\psi^\dagger_{\pm}$, eq.\eqref{boundary}.

Once again, as we did in \eqref{restoran}, we can be more explicit by displaying the full matrix structure in the equation obeyed by the Green function in the $d=1$ case,
\[{\cal D}^0_x\, {\cal G}_{xy}= \delta(t_x-t_y) \delta(x-y) I_{4 \times 4},\]
where
\beq
{\cal D}^0=i\left(
\begin{array}{cccc}
0 & \partial_0 + \partial_1 & 0 & 0\\
\partial_0 - \partial_1 & 0 & 0 & 0\\
0 & 0 & 0 & \partial_0 + \partial_1\\
0 & 0& \partial_0 - \partial_1 & 0\end{array}
\right)
\eeq
and
\beq
{\cal G}=i\left(
\begin{array}{cccc}
0 & {\cal G}^R_{ret} & 0 & {\cal G}^R_{K}\\
{\cal G}^L_{ret}& 0 & {\cal G}^L_{K} & 0\\
0 & 0 & 0 & {\cal G}^R_{adv}\\
0 & 0& {\cal G}^L_{adv} & 0\end{array}
\right)
\eeq
Here we have defined retarded, advanced and Keldysh components of the Green function as
\beq
{\cal G}^i_{ret}&=&G^i_{++}-G^i_{+-}=G^i_{-+}-G^i_{--}\;,  \nonumber\\
{\cal G}^i_{adv}&=&G^i_{++}-G^i_{-+}=G^i_{+-}-G^i_{--}\;,  \;\;\;\;\;\;  i= R,L\nonumber\\
{\cal G}^i_{K}&=&G^i_{++}+G^i_{--}=G^i_{+-}+G^i_{-+} \;.\label{retadvk}
\eeq

\subsection{Incorporating  sources}

Let us now include the source term appearing in \eqref{principio}, which, after the doubling of degrees of freedom, reads
\begin{equation}
\bar{\psi}_a\,s_{ab}\,\psi_b=\bar{\psi}_+ \lns_+\psi_+ -\bar{\psi}_- \lns_-\psi_- ,
\end{equation}
where the vector source will be written as
\[s=\left(\!\!\begin{array}{cc}\lns_+\! & 0\\
0\! & -\lns_-\\
\end{array}\!\!\right).\]
Performing the rotation in fermion fields (\ref{LO}), we obtain
\begin{equation}
\bar{\psi}_a\,s_{ab}\,\psi_b= \bar{\Psi}\,\not \!{\cal S}\,\Psi,
\end{equation}
where the matrix ${\cal S}$ takes the form
\[\not \!{\cal S}=\left(\!\!\begin{array}{cc}\frac{\lns_+ +\lns_-}{2}\! & \frac{\lns_+ -\lns_-}{2}\\
\frac{\lns_+ -\lns_-}{2}\! & \frac{\lns_+ +\lns_-}{2}\\
\end{array}\!\!\right).\]
Note that, in contrast to Dirac operator ${\cal D}^0$,   ${\cal S}$ is a non-diagonal matrix.
At this point, it is convenient to define
\beq \lns_c=\frac{\lns_+ +\lns_-}{2}\,\,\,\,\,\,\,\,\,\,\,\,\,\,\,\,\,\,  \lns_q=\frac{\lns_+ -\lns_-}{2} \!\!,\eeq
where the subscripts $c$ and $q$ stand for ``classical'' and ``quantum'' components of the vector fields (this terminology is habitual in the context of quantum non-equilibrium field theory, it refers to the role played by the fields when studying the classical equations of motion \cite{Das}). Now, the matrix ${\cal S}$ can be written in the form

\beq \label{A}\not\!{\cal S}= \lns_c \lambda^c + \lns_q \lambda^q  \equiv \lns_a\lambda^a\, , \;\; a = c,q
\label{lll}\eeq
with
\[\lambda^c=\left(\!\!\begin{array}{cc}1\! & 0\\
0\! & 1\\
\end{array}\!\!\right),\;\;\;\;\;\lambda^q=\left(\!\!\begin{array}{cc}0\! & 1\\
1\! & 0\\
\end{array}\!\!\right)\]
At this point we have all the necessary ingredients to resume our main task.

\section{Duality and bosonization}

Taking into account that the Jacobian of the change of variables proposed in the previous sections is trivial (i.e. field-independent), the fermionic-path integral in (\ref{partition}) can be written in terms of the fermion fields $\Psi$ and $\bar{\Psi}$ introduced in eq.\,\eqref{LO}. Once the source term is introduced, the same can be done for the generating functional which  in terms of the new variables takes the form
\begin{equation}\label{fer-det2}
Z_F[s_{\mu}] = \int {\cal D}\bar{\Psi}\,{\cal D}\Psi\,e^{i S_F[\Psi,\bar{\Psi},{\cal S}(s)]}
\end{equation}
where
\begin{equation}
S_F = \int dt\,d^{d}x \,\bar \Psi \left({\cal D}^0 + \not\! {\cal S}(s) \right) \Psi.
\label{volvio}
\end{equation}

  Let us now introduce the functional bosonization procedure \cite{modern-bos}. Although this formulation can be in principle used for arbitrary space-time dimensions, in order to make contact with the results of \cite{GGM}, from now on we will restrict our study to $1+1$ dimensions ($d=1$) which is besides the only case in which   bosonization procedure is exact. Let us perform a  local change of path-integral variables

\begin{equation}
\Psi \rightarrow g(x)\,\Psi,
\end{equation}
where $g(x)$ is composed of two successive $U(1)$  transformations
\beq
g(x) = g_c(x) g_q(x)
\eeq
where
\begin{eqnarray}
g_c(x) &=&  \exp(i \alpha_c(x)\lambda^c ) \label{gg1}\\
g_q(x) &=&
 \exp(i \alpha_q(x)\lambda^q ) \label{gg2}
\end{eqnarray}
 {The boundary conditions on $\alpha = (\alpha_c,\alpha_q)$ should be chosen so that those on fermion fields, eqs.\eqref{boundary}
remain valid. To obtain the appropriate boundary conditions for $\alpha$ we shall consider in what follows a general transformation depending  on functions that will become eventually bosonic fields that we shall denote generically as $\phi$.}

 {Let us introduce a notation describing boundary conditions for the original fields $\psi$ when written in terms
of $\psi_\pm$ and $\psi_\tau$
 \begin{eqnarray}
\psi_{b_r}(x, t_i) = \epsilon_{b_s}\psi_{{b_s}}(x,t_i)
\label{linda}
\end{eqnarray}
where index $i$ denotes the different boundary points in the time contour ($t_i = -\infty, \infty, 0,\beta$) and
$\psi_{b_r},\psi_{b_s}$ indicate the fields in different regions of the contour ($b_r,b_ss= +,-,0 \tau$). Concerning $\epsilon_{b_s}$
it takes the value $\epsilon_s = +1$ for $s=\pm, 0 $ and $\epsilon_s = -1$ for $s=\tau$ (see eq.\eqref{boundary}.  }

 {
Under a transformation $U_{b_r}[\phi_{r_i}(x,t)] = U_{b_r}(x,t)$ the fermion fields at the boundary transform according to
\beq
\psi_{b_r}(x, t_i) \to \psi_{b_r}(x, t_i)' = U_{b_r}(x,t_i) \psi_{b_r}(x, t_i)
\eeq
Now, since  $\psi_{b_r}(x, t_i)' $ should obey the same boundary conditions  as $ \psi_{b_r}(x, t_i) $, namely \eqref{linda},
one has that
 \beq
 U_{b_r}(x,t_i) = U_{b_s}(x,t_i)
 \label{28}
\eeq
where there is no $\epsilon_s$ factor as in the fermion case so as to preserve the signs implied by eq.\eqref{linda}}. These boundary conditions are then those satisfied by $g$ and also by all bosonic fields to be introduced below.

One can always define a path-integral measure such that   the Jacobians associated to  transformations \eqref{gg1}-\eqref{gg2} are trivial \cite{Fidel-et-al1}-\cite{Naon} so that
after the change of variables,  the generating functional \eqref{fer-det2} becomes

\beq
Z_F[s_{\mu}] = \int {\cal D}\bar{\Psi}\,{\cal D}\Psi\,\exp\big(i\int d^{2}x \,\bar \Psi \left({\cal D}^0 + \not \!{\cal S}(s)+ig^{-1}\slp g \right) \Psi \big).
\label{NN}
\eeq
We shall now introduce  an auxiliary field ${\cal B}_{\mu}$
\beq
{\cal B}_\mu = {\cal B}_\mu^a \lambda^a \; , \;\;\;\; a=c,q
\eeq
together with a delta-function condition so that it represents a pure gauge,
\beq
\not \! {\cal B} + ig^{-1}\slp g = 0
\label{cero}
\eeq

Now, instead of \eqref{NN}  we write the fermionic generating functional in the form
\beq
Z_F[s_{\mu}] = \int {\cal D}{\cal B}_{\mu}{\cal D}\bar{\Psi}\,{\cal D}\Psi\,\exp\big(i\int d^2x\, \bar \Psi \left({\cal D}^0 + \not \!{\cal S}(s)+ \not \!{\cal B} \right) \Psi \big)\,\,\delta({\cal F}({\cal B})),
\eeq
where
\beq
{\cal F}({\cal B})=\epsilon_{\mu\nu}\partial^{\mu}{\cal B}^{\nu}.
\eeq
imposes the pure gauge condition on ${\cal B}_\mu^a$. As explained before, the bosonic field  ${\cal B}_\mu^a$
obeys   boundary conditions \eqref{28}.

 Taking into account that the fermionic path-integral is just the determinant of the Dirac operator, and the  shift
 \beq
  \not\!{\cal B}+\not\!{\cal S}(s)\rightarrow {\cal B},
  \eeq
 one ends up with
\beq\label{24}
Z_F[s_{\mu}] = \int {\cal D}{\cal B}_{\mu}\det \left({\cal D}^0 + {\cal B} \right)\,\,\delta({\cal F}({\cal B})-{\cal F}({\cal S})).
\eeq
We now introduce in the path-integral defining  $Z$ a scalar field $\Phi = \Phi_a\lambda^ai$ to represent  the delta-function constraint in the form
\beq
\delta({\cal H}) = \int {\cal D}\Phi\,\exp\left(i C\, tr_K \int d^2x\,{\cal H}\,\Phi\right)\,
\label{deltasf}
\eeq
with  $C$  a constant to be appropriately fixed below  $tr_K$ indicating a  trace over matrices $\lambda^a$.
 Being $\Phi$ a bosonic field boundary conditions correspond to those given by eq.\eqref{28}.

Using the delta-function representation \eqref{deltasf} the generating functional $Z_F[s]$ takes the form
\beq\label{25}
Z_F[s] = \int {\cal D}\Phi\,e^{i S_B[\Phi]}\,\exp\left(-i C\, tr_K \int d^2x{\cal F}({\cal S}[s])\,\Phi \right),
\eeq
with  $S_B[\Phi]$ defined as
\beq
e^{i S_B[\Phi]} &=& \int {\cal D}{\cal B}_{\mu}
 \det \left({\cal D}^0 + {\cal B} \right)\,\exp\left(i C tr_K \int d^2x\,{\cal F}({\cal B})\,\Phi \right).
\label{bosonic-action}
\eeq
Given  $Z_F$ written as in \eqref{25} one can interpret $S_B[\Phi]$ as a bosonic action for the scalar field $\Phi$ and then  establish an identity between the original fermionic  generating functional \eqref{fer-det2} and a bosonic generating functional
\beq
Z_B[s] = \equiv \int {\cal D}\Phi\,e^{i S_B[\Phi]}\,\exp\left(-i C\, tr_K \int d^2x{\cal F}({\cal S}[s])\,\Phi \right),
\eeq
That is
\beq
Z_F[s] = Z_B[s]
\label{37}
\eeq

In order to find an explicit form for the bosonic action  $S_B[\Phi]$ it is necessary to compute the fermionic determinant and then proceed to integrate over $b$. Once this is done one should obtain an identity between the original fermionic generating functional and  what one can interpret as a bosonic generating functional $Z_B[\phi]$,
\beq
Z_B[s_\mu] = \int {\cal D}\Phi\,e^{i S_B[\Phi]} \exp\big(-i C tr_K \int d^2x\,{\cal S}_\mu(s) \epsilon_{\mu\nu}\partial_\nu \Phi   \big),
\eeq

Thus, bosonization of the original fermionic theory has been achieved through the identity
\beq\label{27}
Z_F[s_{\mu}] = Z_B[s_{\mu}],
\eeq

This formula  allows to obtain, through functional derivation, the bosonic representation for fermionic currents.

In Appendix \ref{ApA} we present a careful evaluation of the fermionic determinant which takes the form
\beq
\det \left({\cal D}^0 + {\cal B} \right)= \det {\cal D}^0  J_F^{dec}[\varphi] J_F^{dec}[\eta]
\label{cuarenta}
\eeq
where the field ${\cal B}$ has been written in terms of the general decomposition
\beq
{\cal B} = \slp \gamma_5 \varphi + \slp \eta
\eeq
and $J_F^{dec}[\chi]$ is defined as (see Appendix)
\begin{eqnarray}
J^{dec}_F[\chi] &=&\exp{\frac{i}{2\pi}\,\int d^2x\,\big((\partial_0 +\partial_1)\chi_q(\partial_0 -\partial_1)\chi_c   +(\partial_0 +\partial_1)\chi_c (\partial_0 -\partial_1)\chi_q\big)}
\label{jacob-finalTexto}
\end{eqnarray}

Now, inserting this result in \eqref{bosonic-action}
one has
\begin{eqnarray}
e^{i S_B[\Phi]} \!\!\!&=& \!\!\! \det \left({\cal D}^0  \right) \int  {\cal D}\eta J_F^{dec}[\eta] {\cal D}\varphi J_F^{dec}[\varphi]
\,\exp\left(-i C tr_K \int d^2x\, \Phi \partial_\mu\partial^\mu \varphi \right)
\label{bosonic-action2}
\end{eqnarray}
We see that the $\eta$ field is completely decoupled from $\Phi$ so that the integral over $\eta$ just gives a trivial factor. Concerning  the remaining  $\varphi$ integral, apart from the term quadratic in $\varphi$ there is the   linear term that couples $\varphi$ with $\Phi$.
\[
 e^{i S_B[\Phi]} =
 {\cal N}  \int  {\cal D}\varphi  \exp\left(\frac{i}{2\pi}\,\int d^2x\, \left(\vphantom{\frac12}(\partial_0 +\partial_1)\varphi_q(\partial_0 -\partial_1)\varphi_c   +(\partial_0 +\partial_1)\varphi_c (\partial_0 -\partial_1)\varphi_q - 2i C   \, \Phi_q \partial_\mu\partial^\mu\varphi_q +
 \Phi_c \partial_\mu\partial^\mu\varphi_c
 \right) \right)
\]
Then, integrating over $\varphi$  one gets for  $S_B[\Phi]$
\beq
e^{i S_B[\Phi]} = \exp\left(i \pi C^2 tr_K \int d^2x\,\lambda^q\,\partial_{\mu}\Phi\partial^{\mu}\Phi\right)
\eeq
Inserting the result   in the r.h.s. of \eqref{25} we finally obtain
\beq
Z_B[s_{\mu}] ={\cal N} \int {\cal D}\Phi\,\exp\left(i \pi C^2 tr_K \int d^2x\,\lambda^q\,\partial_{\mu}\Phi\partial^{\mu}\Phi\right)\,\exp\left(-i C tr_K \int d^2x\,{\cal S}_{\mu}\,\epsilon^{\mu\nu}\partial_{\nu}\Phi \right),
\eeq
where the factor ${\cal N}$, being independent of the fields and the sources, is a normalization factor that can be disregarded when computing correlation functions. At this point we also note that the constant $C$ does not play any relevant role, as expected. Indeed, it becomes apparent that it can be eliminated from the action by rescaling the field. In order to facilitate comparison with previous results we redefine $2\sqrt{\pi}C\phi\rightarrow\phi$ obtaining
\beq
Z_B[s_{\mu}] ={\cal N} \int {\cal D}\phi\,\exp\left(\frac{i}{4} tr_K \int d^2x\,\lambda^q\,\partial_{\mu}\Phi\partial^{\mu}\Phi\right)\,\exp\left(-\frac{i}{2\sqrt{\pi}} tr_K \int d^2x\,{\cal S}_{\mu}[s]\,\epsilon^{\mu\nu}\partial_{\nu}\Phi \right),
\label{veni}
\eeq
with ${\cal N}$ a normalization constant.

From the  generating functional \eqref{veni} we see that the  bosonic  dual Lagrangian density takes the form
\beq
{\cal L}_B = \frac{1}{4}\,tr_K\,\lambda^q\,\partial_{\mu}\Phi\partial^{\mu}\Phi.
\eeq
This bosonized Lagrangian can be written in a more illuminating way  introducing the boson fields $\phi_+$ and $\phi_-$ such that
\begin{eqnarray}
\phi_+ &=&\frac{\phi^c +\phi^q}{\sqrt{2}} \nonumber\\
\phi_- &=&\frac{\phi^c -\phi^q}{\sqrt{2}}
\end{eqnarray}
In terms of the $\phi_\pm$ fields   ${\cal L}_B$ takes the form
\beq\label{bosonic-Lagrangian}
{\cal L}_B = \frac{1}{2}\,(\partial_{\mu}\phi_{+}\partial^{\mu}\phi_{+}-\partial_{\mu}\phi_{-}\partial^{\mu}\phi_{-}).
\eeq
which corresponds to a free boson model.

{As it was the case for fermions $\psi_\pm$, although the bosonic fields  $\phi_\pm$   in   bosonic  Lagrangian \eqref{bosonic-Lagrangian} semble decoupled, they are in fact correlated through the bosonic time boundary conditions
\beq
\begin{array}{ll}
\phi_{+} (x,t =+\infty) & =  \phi_{-}(x, t= +\infty) \label{boundary2} \\
\phi_{-} (x,t =-\infty) & = \phi_{\tau}(x, t= 0) \\
\phi_{+} (x,t =-\infty) & = \phi_{\tau}(x, t= \beta)
\end{array}
\eeq
}

The action associated to Lagrangian \eqref{bosonic-Lagrangian}
\begin{equation}
S_B[\phi_+,\phi_-]= \frac12 \int dx  \int_{-\infty}^\infty dt
 (\partial_{\mu}\phi_{+}\partial^{\mu}\phi_{+}-\partial_{\mu}\phi_{-}\partial^{\mu}\phi_{-}).
\label{actionBB}
\end{equation}
can be also written in the form
\beq
S_B[\phi] =\frac12 \int_{\cal C} dtdx  \, \partial_\mu \phi \partial^\mu \phi
\label{actuna}
\eeq
where  ${\cal C}$ is the closed time contour introduced  in the original fermionic action \eqref{principio}.

One can compute vacuum expectation values of fermion current correlations in terms of bosonic currents using the identity between  fermionic and bosonic generating functionals established in \eqref{37}. As an example, one has
\begin{eqnarray}
\langle \bar{\psi}_{\pm}(x_1)\,\gamma_{\mu_1}\,\psi_{\pm}(x_n) \ldots
\bar{\psi}_{\pm}(x_n)\,\gamma_{\mu_n}\,\psi_{\pm}(x_n)\rangle_F &=&\left. \frac1{Z_F[s]} \frac{\delta^n Z_F[s]}{\delta s_\pm^{\mu_1}(x_1)\ldots \delta s^{\mu_n}_\pm(x_n)} \right|_{s=0}\nonumber\\
&=&\left. \frac1{Z_B[s]} \frac{\delta^n Z_B[s]}{\delta s_\pm^{\mu_1}(x_1)\ldots \delta s^{\mu_n}_\pm(x_n)} \right|_{s=0}\nonumber\\
&=&\left(\mp (2\pi)^{-1/2}\right)^n \langle\epsilon _{\mu_1\nu_1} \ldots
\epsilon _{\mu_n\nu_n} \partial_{\nu_1}\!\phi_\pm(x_1) \ldots \partial_{\nu_n}\!\phi_\pm(x_n)\rangle_B
\end{eqnarray}
We can summarize our bosonization results in terms of the following fermion-boson correspondence
\begin{eqnarray}
{\cal L}_F = \bar{\psi}_{+} i \slp \psi_+ -  \bar{\psi}_{-} i \slp  \psi_{-}
&\rightarrow& {\cal L}_B = \frac{1}{2}\,(\partial_{\mu}\phi_{+}\partial^{\mu}\phi_{+}-\partial_{\mu}\phi_{-}\partial^{\mu}\phi_{-}), \label{rules1}\\
\bar{\psi}_{+}\,\gamma^{\mu}\,\psi_{+}&\rightarrow &-\frac{1}{\sqrt{2\pi}}\,\epsilon^{\mu\nu}\partial_{\nu}\phi_{+}, \label{rules1.5}
\\
\bar{\psi}_{-}\,\gamma^{\mu}\,\psi_{-} &\rightarrow&\frac{1}{\sqrt{2\pi}}\,\epsilon^{\mu\nu}\partial_{\nu}\phi_{-} \, .
\label{rules2}
\end{eqnarray}
In view of the the connection \eqref{rules1}, one can write a bosonization recipe for  the action written in terms of the Keldysh contour ${\cal C}$ in the form
\begin{equation}
S_F[\psi,\bar{\psi}]= \int_{\cal C}\,dt\,dx \, \bar{\psi} i \slp \psi  \rightarrow
S_B[\phi]= \frac{1}{2}\int_{\cal C}\,dt\,dx \, \partial_\mu\phi \,\partial^\mu\phi
\label{principio2}
\end{equation}
To write this formula we have used the fact that fields on the  ${C}_\tau$ contour are decoupled from sources thus playing no role in final expressions.

Eq.\eqref{principio2} is formally identical to the Coleman-Luther-Mandelstam-Mattis bosonization recipe in the equilibrium case \cite{Stone}.
Concerning the  bosonization recipe for currents eqs.\eqref{rules1.5}-\eqref{rules2}
note that they both lead to the same conserved charge, $dQ/dt = 0$ since the relative sign of the currents is compensated by the opposite sign of the time-integral limit.
\beq
Q = \int dx^1 \bar\psi \gamma^0 \psi \rightarrow \int dx^1 \phi \,\partial_1 \phi
\eeq

 {Let us conclude by stressing that we were able to arrive to the simple bosonization rules \eqref{rules1}-\eqref{rules2} implying a quadratic bosonic Lagrangian because one can compute the fermion determinant
$ \det \left({\cal D}^0 + {\cal B} \right)$  in a closed form, this being a key tool in our approach.  This was the reason why  we have considered a Lagrangian for both positive and negative  chirality fermions  (cf with \cite{GGM}) since, at it is well known  \cite{AGG} if just one chirality is considered,  the Dirac operator does not lead to a well-defined eigenvalue problem and hence the definition of an associated Dirac determinant  is not a  well-defined problem.}

\section{Current-current correlations and the connection between fermionic and bosonic out of equilibrium distribution functions}
Here we shall evaluate vacuum expectation values of current correlations, by using the bosonization recipe \eqref{rules1}-\eqref{rules2}. More specifically we shall consider the following correlation function
\beq\label{current-current}
<(\bar{\psi}_{+}\,\gamma^{\mu}\,\psi_{+}(x))\,(\bar{\psi}_{-}\,\gamma^{\nu}\,\psi_{-})(y)>_F =
 -\frac{1}{2\pi }\,<\epsilon^{\mu\rho}\partial_{\rho}\phi_{+}(x)\,\epsilon^{\nu\sigma}\partial_{\sigma}\phi_{-}(y)>_B,
\eeq
where the l.h.s. is to be computed for the original free fermionic model, and the r.h.s. with the bosonic action for Lagrangian \eqref{bosonic-Lagrangian}.
The Fourier transform of the out of equilibrium right (R) and  Left (L)  fermionic Green functions \eqref{retadvk} are given by
\[
{\cal G}^R_{ret}(k)= \frac{1}{k^0 - k^1 + i \delta} \, , \;\;\;\;\; {\cal G}^L_{ret}(k)= \frac{1}{k^0 +  k^1 + i \delta}
\]
\[
{\cal G}^R_{adv}(k)= \frac{1}{k^0 - k^1 - i \delta} \, , \;\;\;\;\; {\cal G}^L_{adv}(k)= \frac{1}{k^0 +  k^1 - i \delta}
\]
\[
{\cal G}^R_{K}(k)= F(k^0)({\cal G}^R_{ret}(k)-{\cal G}^R_{adv}(k))\, , \;\;\;\;\; {\cal G}^L_{K}(k)= F(k^0)({\cal G}^L_{ret}(k)-{\cal G}^L_{adv}(k)),
\]
where $F(k^0)$ is the fermionic Wigner function, connected to the generic (not necessarily the one corresponding to equilibrium) fermionic distribution function $n_{F}(k^0)$
\beq
F(k^0)=1-2 n_{F}(k^0)
 \eeq
 At thermodynamical equilibrium $n_{F}(k^0)$ coincides with the Fermi-Dirac distribution and then one has
 \[F^{eq}(k^0)= \tanh(\frac{\beta k^0}{2})\]
 where $\beta=1/kT$.

The free bosonic Green functions  arising  from the dual bosonic Lagrangian \eqref{bosonic-Lagrangian}, with out of equilibrium bosonic distribution $n_{B}(k^0)$ can be expressed in terms of the inverse of the d'Alembertian  operator  $\Delta$ as
\[\Delta_{ret}(k)= \frac{1}{k_0^2 + k^1k_1 + i \delta}\]
\[\Delta_{adv}(k)= \frac{1}{k_0^2  + k^1k_1 - i \delta}\]
\beq
\Delta_{K}(k)= B(k^0)(\Delta_{ret}(k)-\Delta_{adv}(k)),
\label{listedabove}
\eeq
where
\beq
B(k^0)=1+2n_{B}(k^0)
 \eeq
 is the bosonic Wigner function. At equilibrium, $n_{B}(k^0)$ is the Bose-Einstein distribution and one has
 \[
 B^{eq}(k^0)=\coth(\frac{\beta k^0}{2})
 \]

We shall consider for definiteness the vacuum expectation values \eqref{current-current} for $\mu=\nu=1$. Using
\beq
<\phi_{+}(x)\,\phi_{-}(y)>_B=2i\Delta_{+-}(x-y)
\eeq
 in the bosonic side we get
\beq
G^R_{+-}(z)G^R_{-+}(-z)+ G^L_{+-}(z)G^L_{-+}(-z)=\frac{i}{\pi}\,\partial_{0}^x\,\partial_{0}^y \Delta_{+-}(z)
\eeq
with $z=x-y$.

In terms of the  physical Green functions \eqref{listedabove}   $\Delta_{+-}$ takes the form
\beq
\Delta_{+-}=\frac{1}{2}(\Delta_{K}+\Delta_{adv}-\Delta_{ret}) \;
\eeq
Then working in momentum space and using the explicit form of   Green functions we get
\beq
\int_{-\infty}^{+\infty}\!\!\!\!dk_0 \int_{-\infty}^{+\infty}\!\!\!\!dk'_0 n_{F}(k^0)(1-n_{F}(k'^0)) (e^{-i(k_0-k'_0)u}+e^{-i(k_0-k'_0)v})=\frac{1}{2}\int_{-\infty}^{+\infty}\!\!\!\!dk_0\,\abs{k_0} \,(B(k^0)-1)(e^{-ik_0u}+e^{-ik_0v})
\, ,
\eeq
where $u=z^0+z^1$ and $v=z^0-z^1$. Multiplying both sides by $e^{i\omega u}$($e^{i\omega v}$) and integrating in $u$ ($v$), we find
\beq
n_B(\omega)= \frac{1}{\omega}\int_{-\infty}^{+\infty}dk_0 n_{F}(k^0)\,[1-n_{F}(k^0-\omega)].
\label{59}
\eeq
This is a remarkable equation that establishes the connection between the fermionic and bosonic distributions for a generic, nonequilibrium situation.

As a check, if one considers the case of equilibrium in which
\[
n^{eq}_{F}(k_0)=1/(e^{\beta k_{0}} +1)
\]
one can easily see that formula \eqref{59} leads to the correct
equilibrium  Bose-Einstein distribution function
\[
n^{eq}_{B}(k_0)=1/(e^{\beta k_0}-1)
\].

\section{Summary and discussion}
The path-integral approach originally developed for bosonization in $D=2$ space-time dimensions as a particular case of a duality transformation \cite{Betal}-\cite{modern-bos} has been very successful  particularly because it allows to simple extensions to $D>2$ dimensions. The duality process amounts to gauging the global symmetry of the original (fermionic) theory, and constraining the corresponding field strength to vanish by introducing a Lagrange multiplier that is finally identified as the dual bosonic field. In this way one can find for example that in the large mass limit of a $D=3$ fermionic theory bosonization leads to a Chern-Simons model, this leading to  the existence
of operators of the Fermi theory which ought to exhibit fractional statistics \cite{modern-bos}. Also in the case of fermions coupled to gravity bosonization can be achieved and in this cases the parity breaking that takes place in $D=3$ Dirac fermion models can be exploited to simulate the effects of crystal defects on the electronic degrees of freedom of topological insulators in condensed matter physics
\cite{Fr1}-\cite{FMoS}.

We have presented in this work an extension of the above referred duality approach to bosonization to the case of out of equilibrium fermion models based in the Schwinger-Keldish formalism. In this way starting from the generating functional $Z_F[s] $ for $D=2$ Dirac fermions coupled to an external source $s_\mu$  written in terms of a path-integral in which the action $S_F$ introduced in \eqref{principio}  is defined in a close time contour we were able to show the identity of $Z_F[s]$ with $Z_B[s]$ , bosonic   generating functional in which the action \eqref{actuna} is also defined in the closed time contour. As in the fermionic case, the apparently decoupled boson field defined in each branch of the time contour are in fact connected through the boundary conditions \eqref{boundary2}.

Concerning the bosonization rules for currents, one finds in each branch of the closed time contour the usual result corresponding to a topologically conserved current $j^\mu_\pm \propto \epsilon^{\mu\nu}\partial \phi_\pm$ whose spatial integral coincides with the Pontryagin class,  this showing the connection of  bosonization with the existence of quantum anomalies, in  the present case with the $D=2$ chiral anomaly which arises through the nontriviality of the Fujikawa Jacobian \cite{Fujikawa}.

By studying current-current correlations  within our bosonization approach we have been able to find in a very simple way the relation between the fermionic $n_F$ and bosonic $n_B$ distribution functions for a generic  non equilibrium situation, eq.\eqref{59}. This relation  agrees with the results of ref.\,\cite{GGM}. The connection
leads to the correct  Bose-Einstein distribution function in equilibrium if $n_F$ is taken as the Fermi-Dirac distribution function.

Although we have considered the free Dirac fermion Lagrangian coupled to an external source, as in the equilibrium case the bosonization rules \eqref{rules1}-\eqref{principio2} are universal in the sense that they hold for interacting fermion models \cite{Stone}. Indeed,  as first observed in the case of the massless and massive self-interacting Luttinger/Thirring model where bosonization was originally derived within the operator approach \cite{Coleman}-\cite{Mandelstam},\cite{M1}-\cite{M2}, the bosonization rules for fermion currents are independent of the coupling constants of the theory thus coinciding with those for free fermions. Hence, a current-current self-interaction as in the Luttinger/Thirring model can be automatically bosonized using the free fermion recipe. This can be easily understood  in the path-integral framework where the current-current interaction can be traded to a linear one using a Hubbard-Stratonovich transformation
\cite{Fidel-et-al3},\cite{Naon}. The same holds in the case of fermions coupled to Abelian and non-Abelian gauge theories \cite{Fidel-et-al1},\cite{Fidel-et-al2}-\cite{Fidel-et-al33}.  We plan for future work the analysis of these models in the out of equilibrium case.

 As stated in the introduction, out of equilibrium bosonization
was first developed in ref.~ \cite{GGM} in a functional integral approach that differs from ours in the way the source terms are handled. The main difference lies in the fact that using the duality technique that we employed here,  the external source can be factored out from the path-integral that allows   to   calculate  the bosonic action in a closed form
and only remains linearly coupled to the dual boson field leading to the correct bosonized recipe. In spite of this difference, both approaches
lead to the same  relation between distribution functions.

We already mentioned that the duality approach to bosonization of fermions in equilibrium can be easily extended to $D>2$ space-time dimensions so that   the analysis presented here for the out of equilibrium case   could be applied with no apparent problems to higher dimensions. Of particular relevance for condensed matter problems is the $D=3$ case in connection with fractional statistics and when coupled to a gravitational background, in applications to study the problem of topological insulators.

 {The bosonization approach that we presented is based on duality transformations related to the global $U(1)$ symmetry of the original fermion theory. Since duality transformations can be easily extended to the  non-Abelian case, one could develop  without much effort an out of equilibrium non-Abelian bosonization along the lines of ref. \cite{modern-bos}.}

We expect to come back to all these issues in a future work.

\vspace{.5cm}

\noindent\underline{Acknowledgments}:   This work was supported by  CONICET  , ANPCYT , CIC  and UNLP, Argentina.

\appendix
\section{Computation of the fermionic determinant:
The Fujikawa Jacobian in Schwinger-Keldysh framework \label{ApA}}
In order to compute the fermionic determinant
\beq
\det ({\cal D}^0 + {\cal B}) = \int {\cal D}\bar\Psi \int {\cal D}\Psi
\exp\big(i\int d^2x\, \bar \Psi \left({\cal D}^0 +  \not \!{\cal B} \right) \Psi \big)
 \eeq
we follow   \cite{Fidel-et-al1} and  perform  a change in the path-integral fermionic variables
\beq\label{change-of-var}\Psi=U \Psi' \,\,\,\,\,\,\,\,\,\,\,\,\,\,\,\,\,\,  \bar{\Psi}=\bar{\Psi'}\bar{U}, \!\!
 \eeq
where $U$ and $\bar{U}$ are given by
\beq U=\exp{i \gamma_5 \varphi} \,\,\,\,\,\,\,\,\,\,\,\,\,\,\,\,\,\,  \bar{U}=\exp{i \gamma_5 \varphi}.
\label{change}\!\!\eeq
In these expressions $\varphi$ is a scalar field that belongs to the algebra generated by the $\lambda^a$ matrices ($\varphi=\varphi_a \lambda^a$) and $\gamma_5 = \gamma_0 \gamma_1$, with the two-dimensional Dirac matrices
\[\gamma^{0}=\left(\!\!\begin{array}{cc}0\! & 1\\
1\! & 0\\
\end{array}\!\!\right),\;\;\;\;\;\gamma^{1}=\left(\!\!\begin{array}{cc}0\! & 1\\
-1\! & 0\\
\end{array}\!\!\right).\]

 {As explained  in section III scalar fields like  $\varphi$  defined in the Keldysh closed time contour    obey   the following should obey the boundary conditions (see eqs.\eqref{boundary2})
\begin{eqnarray}
\varphi_{+} (x,t =+\infty) &=&  \varphi_{-}(x, t= +\infty)\nonumber\\
\varphi_{-} (x,t =-\infty) &=& \varphi_{\tau}(x, t= 0)\nonumber\\
\varphi_{+} (x,t =-\infty) &=& \varphi_{\tau}(x, t= \beta)\nonumber\\
\label{boundary3}
\end{eqnarray}
}

Now one can always choose a gauge (the ``decoupling gauge'' introduced in ref.\cite{Fidel-et-al2}) in which  ${\cal B}$ can be written  in terms of a   scalar field  $\varphi$ in in the form
\begin{equation}\label{A1}
{\cal B}^{dec}=-i\,\bar{U}\,\slp U = \slp \gamma_5 \varphi,
\end{equation}
 {Boundary conditions for ${\cal B}^{dec}$ coincide with those for $\varphi$ given in eq.\eqref{boundary3}}

After the change of variables \eqref{change-of-var}-\eqref{change}  one then ends with
\beq
\det \left({\cal D}^0 + {\cal B}^{dec} \right) = J_F^{dec}[\varphi] \det {\cal D}^0
\label{26}
\eeq
where $ J_F^{dec}[\varphi]$ is the Fujikawa Jacobian \cite{Fujikawa} associated to the change of variables
\begin{equation}\label{measure}
{\cal D}\bar{\Psi}\,{\cal D}\Psi=J_F^{dec}\,{\cal D}\bar{\Psi'}\,{\cal D}\Psi'.
\end{equation}
Note that the  ${\cal D}^0$ determinant resulting from the $\bar \Psi', \Psi$ path-integration is a constant independent of ${\cal B}$.
It should be stressed that Fujikawa's method implies a gauge-invariant regularization of the determinant so that the answer obtained in the decoupling gauge is independent of the gauge choice.

In the case of thermodynamical equilibrium, without the introduction of the closed time contour, $J_F$ has been carefully evaluated \cite{Fidel-et-al2}-\cite{Fidel-et-al33}. Interestingly, we shall see  that the main computational scheme remains valid when considering the closed time path.

The presence of Jacobian $J_F$ is due to the non-invariance of the path-integral measure under the chiral factor $\exp(i\gamma_5 \varphi)$ in the change of variables (\ref{change-of-var}). Even in the context of equilibrium QFT's, { where, usually,  only the forward time path is considered}, the fermionic Jacobian $J_F$  is directly connected with the chiral anomaly, as Fujikawa originally showed \cite{Fujikawa}. Its non-triviality is due to the necessity, in its  computation, of a proper regularization that depends on the symmetries to be preserved on physical grounds \cite{Cabra-Schaposnik}.
This Jacobian  is at the heart of the path-integral approach to fermion bosonization in two dimensional space-time fermion theories \cite{Stone}.

Let us consider an extended transformation $U_{\tau}$, depending on a parameter $\tau$ ($\tau \in [0,1]$):
\beq U_{\tau} =\exp{i\tau \gamma_5 \varphi}.\eeq
and define a $\tau$-dependent operator ${\cal D}^{\tau}$ as
\begin{equation}
{\cal D}^{\tau}={\cal D}^0 + (1-\tau)\slp\gamma_5 \varphi,
\end{equation}
such that for $\tau=0$ it coincides with the original operator ${\cal D}^0 + {\cal B}$ and for $\tau=1$ becomes the free operator. Next we  introduce the quantity
\begin{equation}
W(\tau)=\log{\det{\cal D}^{\tau}},
\label{30}
\end{equation}
such that
\begin{equation}
 -\int_0^1 \,\frac{d W(\tau)}{d\tau}\,d\tau = \log {\cal D}^\tau - \log{\cal D}^0.
 \label{31}
\end{equation} Then, in view of eqs.\eqref{26}-\eqref{31} we have
\beq\label{jacobW}
J_F^{dec} = \exp\left( -\int_0^1 \,\frac{d W(\tau)}{d\tau}\,d\tau \right).
\eeq
Once this Jacobian is calculated,   eq.\eqref{26}
leads to an explicit form for  the fermion determinant .

Now from eq.\eqref{30}, the integrand in the exponential argument can be written as
\begin{equation}\label{W-derivative}
\frac{d W(\tau)}{d\tau}= \frac{d\log{\det{\cal D}^{\tau}}}{d\tau}=Tr({\cal D}^{\tau})^{-1}\frac{d{\cal D}^{\tau}}{d\tau},
\end{equation}
where $Tr$ denotes a trace over Lorentz, coordinate space and also over matrices $\lambda^a$. The calculation of $ {d W(\tau)}/{d\tau}$ involves a divergence coming from the evaluation of the fermionic Green function
 $({\cal D}^{\tau})^{-1}(x,y)$ at the same space-time point.
To see this we write the r.h.s. of (\ref{W-derivative}) in a more explicit way:
\begin{equation}\label{W-derivative-explicit}
\frac{d W(\tau)}{d\tau}= \lim_{y \to x}\,tr_L \otimes tr_K\,\int d^2y\,B(x,y,\tau),
\end{equation}
where $B(x,y,\tau)=[{\cal D}^{\tau}]^{-1}(x,y)\,\slp_y\gamma_5 \varphi(y)$.
It is convenient to specify the equation obeyed by the Green function $[{\cal D}^{\tau}]^{-1}(x,y)$. In Lorentz space, both ${\cal D}^{\tau}$ and $[{\cal D}^{\tau}]^{-1}$ have anti-diagonal matrix structure:
\[{\cal D}^{\tau} [{\cal D}^{\tau}]^{-1}=\left(\!\!\begin{array}{cc}0\! & {\cal D}^{\tau}_{L,x}\\
{\cal D}^{\tau}_{R,x}\! & 0\\
\end{array}\!\!\right)\,\left(\!\!\begin{array}{cc}0\! & {\cal G}_R(x,y,\tau)\\
{\cal G}_L(x,y,\tau)\! & 0\\
\end{array}\!\!\right)=\delta^2(x-y)\left(\!\!\begin{array}{cc}1\! & 0\\
0\! & 1\\
\end{array}\!\!\right),\]
where the subscripts $L$ and $R$ indicate the chiral (left and right) components of each fermion field. Of course, as explained before, ${\cal G}_R$ and ${\cal G}_L$ have, in turn, a triangular matrix structure in Keldysh space (see Section II). On the other hand, $[{\cal D}^{\tau}]^{-1}(x,y)$ is related to the free Green function as
\[[{\cal D}^{\tau}]^{-1}(x,y)\equiv{\cal G}(x,y)=\exp{i(1-\tau)\gamma_5 \varphi(x)}\,{\cal G}^0(x,y)\,\exp{i(1-\tau)\gamma_5 \varphi(y)}
,\]
with
\[{\cal G}^0(x,y)=\left(\!\!\begin{array}{cc}0\! & {\cal G}^0_R(x,y)\\
{\cal G}^0_L(x,y)\! & 0\\
\end{array}\!\!\right)
.\]
Performing the corresponding matrix products (in Lorentz space), one can see that the integrand in (\ref{W-derivative-explicit}) is a diagonal matrix:
\[B(x,y,\tau)=\left(\!\!\begin{array}{cc}B_R(x,y,\tau)\! & 0\\
0\! & B_L(x,y,\tau)\\
\end{array}\!\!\right),\]
where
\[B_{R,L}(x,y,\tau)=\pm \,e^{\pm i(1-\tau)\varphi(x)}\,{\cal G}^0_{R,L}(x,y)\,e^{\mp i(1-\tau)\varphi(y)}(\partial_0\mp\partial_1)\varphi(y).\]
where the upper (lower) sign corresponding to $B_R$ ($B_L$).

After  performing the Lorentz trace  (\ref{W-derivative-explicit}) then reads
\beq\label{W-derivative-explicit2}
\frac{d W(\tau)}{d\tau}= \lim_{y \to x}\,tr_K\,\int d^2y\,\left(B_R(x,y,\tau)+B_L(x,y,\tau)\right).
\eeq
Here we note that, as anticipated, $\lim_{y \to x}\,B_{R,L}(x,y,\tau)$ is divergent. Indeed, in this limit the fermion functions ${\cal G}_{R,L}(x,x)$ represent products of fermion fields at the same point. We shall adopt a point-splitting regularization which amounts to introduce  a   two-vector $\epsilon$ and define the regularized Green functions through a {\em symmetric} $\epsilon \rightarrow 0$ limit:
 \beq
\lim_{y \to x}\,{\cal G}_{R,L}(x,y)\equiv\lim_{\epsilon \to 0}\,\frac{{\cal G}_{R,L}(x,x+\epsilon)+{\cal G}_{R,L}(x,x-\epsilon)}{2},
\eeq
with
\beq
\lim_{\epsilon \to 0}\,\frac{\epsilon^{\mu}\epsilon^{\nu}}{\epsilon^{\rho}\epsilon_{\rho}}=\frac{g^{\mu\nu}}{2},
\eeq
and
\beq
\lim_{\epsilon \to 0}\,\frac{\epsilon^{\mu}}{\epsilon^{\rho}\epsilon_{\rho}}=0,
\eeq
where $g^{\mu\nu}$ is the metric tensor.
This leads to the following expressions:
\beq\label{B}
\lim_{y \to x}\,B_{R,L}(x,y,\tau)=\pm \lim_{\epsilon \to 0}( e^{\pm i(1-\tau)\varphi(x)}\,\frac{{\cal G}^0_{R,L}(x,x+\epsilon)+{\cal G}^0_{R,L}(x,x-\epsilon)}{2} \,e^{\mp i(1-\tau)\varphi(x)}\,\mp \nonumber\\
e^{\pm i(1-\tau)\varphi(x)}\,\frac{{\cal G}^0_{R,L}(x,x+\epsilon)-{\cal G}^0_{R,L}(x,x-\epsilon)}{2} \,e^{\mp i(1-\tau)\varphi (x)}\,\epsilon^{\mu}\partial_{\mu}\varphi(x))(\partial_0\mp\partial_1)\varphi(y).
\eeq
At this point, in order to go further, we must recall that we have to perform the product of the $\lambda^a$ in $\varphi = \varphi_a\lambda^a$ and in the triangular Green function matrices
\beq{\cal G}^0_{R,L}=\left(\!\!\begin{array}{cc}{\cal G}^0_{R,L,ret}\! & {\cal G}^0_{R,L,K}\\
0\! & {\cal G}^0_{R,L,adv}\\
\end{array}\!\!\right),\label{triangular2}.\eeq
Now we have to specify the precise form of the Green functions. For simplicity we will use an equilibrium distribution at zero-temperature. Interestingly, this choice does not imply any loss of generality, since we have confirmed that the regularization performed with Green functions corresponding to a generic out of equilibrium distribution yields the same result for $J_F$. Then we employ the functions
\beq
{\cal G}^0_{R,L,ret}(z,0)=\frac{\theta(z_0)}{2\pi} (\frac{1}{z^0\mp z^1 +i\delta}-\frac{1}{z^0 \mp z^1 -i\delta}),
\eeq
\beq
{\cal G}^0_{R,L,adv}(z,0)=\frac{-\theta(-z_0)}{2\pi} (\frac{1}{z^0\mp z^1 +i\delta}-\frac{1}{z^0 \mp z^1 -i\delta}),
\eeq
\beq\label{kel}
{\cal G}^0_{R,L,K}(z,0)=\frac{-1}{2\pi} \frac{1}{z^0\mp z^1 +i\delta},
\eeq
with $\delta > 0$.

 We find that the symmetric $\epsilon \rightarrow 0$ limit of $\frac{{\cal G}^0_{R,L}(x,x+\epsilon)+{\cal G}^0_{R,L}(x,x-\epsilon)}{2} \epsilon^{\mu}$ vanishes. Concerning $\frac{{\cal G}^0_{R,L}(x,x+\epsilon)-{\cal G}^0_{R,L}(x,x-\epsilon)}{2}$, only its Keldysh component will give a non-vanishing contribution, when contracted with the $\epsilon^{\mu}$ factor, in the second term of (\ref{B}):
\beq\label{kelsylim}
\lim_{\epsilon \to 0}\frac{{\cal G}^0_{R,L}(x,x+\epsilon)-{\cal G}^0_{R,L}(x,x-\epsilon)}{2}\,\epsilon^{\mu}=\frac{g^{0\mu}\mp g^{1\mu} }{4\pi}\left(\!\!\begin{array}{cc}0\! & 1\\
0\! & 0\\
\end{array}\!\!\right),
\eeq
where here $g^{\mu \nu}$ is the metric tensor.

 {This result is valid for an arbitrary nonequilibrium $n_F(k_0)$. To see this we concentrate  in the only component of the Green function that survives the symmetric limit,  the Keldysh one. For simplicity we consider the $R$ component which in the general situation takes the form (compare with \eqref{kel})
\beq\label{kelnoneq}
{\cal G}^0_{R,K}(v,0)=\frac{-1}{2\pi}
\lim_{a \to 0} \int_0^{\infty}dk^0 e^{-a k^0}(1-2n_F(k_0))\sin{k^0 v},
\eeq
Here $v=z^0 - z^1$ is a light-cone coordinate and $a$ an ultraviolet regulator. It is straightforward to get
\beq\label{kelnoneq2}
{\cal G}^0_{R,K}(v,0)=\frac{-1}{2\pi v} + O(v),
\eeq
which, when used in the l.h.s. of \eqref{kelsylim} yields the same symmetric limit as in the equilibrium case.}

Coming back to $d W(\tau)/d\tau$, we can now compute the regularized matrix products in (\ref{B}) and take the Keldysh trace in (\ref{W-derivative-explicit2}), obtaining
\beq
\label{W-derivative-explicit3}
\frac{d W(\tau)}{d\tau}= \frac{i(\tau -1)}{\pi}\,\int d^2x\,\big((\partial_0 +\partial_1)\varphi_q(\partial_0 -\partial_1)\varphi_c+ \nonumber\\ +(\partial_0 +\partial_1)\varphi_c (\partial_0 -\partial_1)\varphi_q \big).
\eeq
Replacing in \eqref{jacobW} and integrating over $\tau$ we get
\begin{eqnarray}
J^{dec}_F[\varphi] =\exp{\frac{i}{2\pi}\,\int d^2x\,\big((\partial_0 +\partial_1)\varphi_q(\partial_0 -\partial_1)\varphi_c   +(\partial_0 +\partial_1)\varphi_c (\partial_0 -\partial_1)\varphi_q\big)}
\label{jacob-final}
\end{eqnarray}
or, writing the Jacobian in terms of ${\cal B^d[\varphi]}$
\beq
J_F^{dec} =
\exp{\left(-i\frac{1}{4\pi}\,tr_K\int d^2x\,{\cal B}^{dec}_{\mu}\,\lambda^q\,{{\cal B}^{dec}}^{\mu}\right)}.
\eeq
 This result has been obtained in the decoupling gauge in which $B_\mu$ was written in the form \eqref{A1}. Going from this gauge to an arbitrary one amounts to write
\begin{equation}\label{A1general}
{\cal B}^{dec}  \to {\cal B}=-i\,\bar{U}\,\slp U  + \slp \eta = \slp \gamma_5 \varphi + \slp \eta
\end{equation}
where the scalar field $\eta$ satisfies the same boundary conditions as $\varphi$ and so does ${\cal B}$. Then     instead of \eqref{jacob-final} one has, in an arbitrtary general case
\beq
J_F[\varphi,\eta] = J_F^{dec}[\varphi]    \times J_F^{dec}[\eta] = \exp{\left(-i\frac{1}{4\pi}\,tr_K\int d^2x\,{\cal B}_{\mu}\,\lambda^q\,{{\cal B}}^{\mu}\right)}.
\label{jacob-finalG}
\eeq

\end{document}